# Generation of electron vortex beams with over 1000 orbital angular momentum quanta using a tuneable electrostatic spiral phase plate

A. H. Tavabi[1*], P. Rosi[2*], A. Roncaglia[3], E. Rotunno[2], M. Beleggia[4], P-H Lu[1], L. Belsito[3], G. Pozzi[4], S. Frabboni[2,4], P. Tiemeijer[5], R. E. Dunin.Borkowski[1], V. Grillo[2]

1. Ernst Ruska-Centre for Microscopy and Spectroscopy with Electrons and Peter Grünberg Institute, Forschungszentrum Jülich, 52425 Jülich, Germany
2. CNR-Nanoscience Institute, S3 center, 41125 Modena, Italy
3. CNR-Institute for Microelectronics and Microsystems, 40129 Bologna, Italy
4. FIM Department, University of Modena and Reggio Emilia, 41125 Modena, Italy
5. Thermo Fisher Scientific, PO Box 80066, 5600 KA Eindhoven, The Netherlands

* Equal contributions

**Abstract:** We report the use of an electrostatic MEMS-based device to produce high quality electron vortex beams with more than 1000 quanta of orbital angular momentum (OAM). Diffraction and off-axis electron holography experiments are used to show that the diameter of the vortex in the diffraction plane increases linearly with OAM, thereby allowing the angular momentum content of the vortex to be calibrated. The realization of electron vortex beams with even larger values of OAM is currently limited by the breakdown voltage of the device. Potential solutions to overcome this problem are discussed.

Scientific breakthroughs often happen simultaneously. In 2010 and 2011, papers from several groups described electron vortex beams [1–4], marking the beginning of a revolution that has made electron microscopists aware of the benefits of measuring and controlling the orbital angular momentum (OAM) of the electron beam in an electron microscope [5–10].

Electron vortex beams are characterised by swirling phase distributions in the azimuthal direction. They are eigenstates of the OAM operator projected onto the electron propagation direction. The number of complete phase turns, $\ell$, is an integer that defines the OAM eigenstate. The expectation of the OAM on these states is $<L> = \ell\hbar$, where $\hbar$ is the reduced Planck constant. Unfortunately, the combination of lenses and multipoles that is available in a modern electron microscope is only sufficient to generate an approximate electron vortex beam, in part because of the requirement for an abrupt potential discontinuity [11]. Instead, the electron beam has to be shaped using purposely-designed electron optical elements.

Based on analogies with light optics, research initially focused on the use of synthetic holograms, *i.e.*, spatially patterned membranes that act on the amplitude and/or phase of the electron beam to produce a desired waveform [12–14]. Although this approach yielded innovative beams [15–18], the resulting electron optical beam-shaping devices suffer from limitations when compared with their counterparts in light optics, including a lack of tunability. As a result, micro-electromechanical systems (MEMS) technology has recently been used to fabricate tuneable phase plates that can be introduced in place of microscope apertures and controlled from outside the electron microscope [19,20]. One interesting idea is to fabricate a programmable phase plate from an array of Einzel lenses to create a discrete approximation of an arbitrary phase landscape [21,22]. Another interesting application of MEMS phase plate technology is the realization of an "OAM sorter", which is also not possible using standard electron optical elements [19].

MEMS devices that have relatively simple architectures can be used to create tuneable vortex beams. One such device, which is referred to as a "chopsticks" device, takes the form of two parallel electrodes that are separated by a narrow gap and have electrical bias voltages applied to them [23,24]. Since the charge distribution on the electrodes is similar to that on a series of parallel dipoles [24], analogies can be drawn with the Aharonov-Bohm effect and with the use of an axially-magnetize needle [25]. As explained in a recent paper [26], every magnetic effect on an electron beam can be reproduced using a set of electrodes. The advantages of using electrostatic elements over magnetic materials include their greater flexibility and tunability, as well as the possibility to use highly compact electrostatic MEMS-based phase plates to introduce comparatively larger phase effects.

Here, we report the experimental realization of very large electron vortex beams, which are (to the best of our knowledge) the largest isolated electron vortex beams (without the additional diffraction orders that are produced by synthetic holograms) that have been created to date. Interest in the generation of such beams has been discussed in a recent paper that demonstrated an $\ell$ = 1000 beam. [27] Potential applications range from magnetic transition radiation [28] to off-axis magnetic field measurement [29,30]. A recent paper by Krenn and Zeilinger [31] raised further interest in the use of such beams to achieve non-paraxial phenomena, such as spin-orbit effects, without the need to increase the numerical aperture/convergence angle. Whereas spin-orbit coupling tends to disappear in the paraxial approximation, Ref. 31 proposes that it should increase with OAM, as it can is proportional to $\bar{L} \cdot \bar{S}$. In the light optical domain, the largest OAM value that has been demonstrated is $\ell$ ~ 10000 [32]. For electrons and synthetic holograms, the record value is currently $\ell$ ~ 1000 in the first diffraction order of the hologram [27]. A larger value of $\ell$ ~ 4000 was achieved in a weak higher diffraction order. However, it was then partially superimposed on other diffraction orders [8].

Here, a large-OAM vortex beam is realised by using an electrostatic device that introduces no absorption or parasitic intensity loss. Furthermore, loss of coherence is limited to the small region where the electrodes are placed and is therefore minor or negligible [22,33].

A scanning electron microscopy (SEM) image of the MEMS-based device, which is referred to here as "MINEON" (MINiaturised Electron Optics for Nanobeams), is shown in Fig. 1. The device is fabricated using silicon-on-insulator technology with trench insulation. The most elevated parts are made from conducting Si that is insulated from the substrate. In this way, the two-dimensional structure defines the conduction path. The device comprises a series of contacts and electrodes around a central hole that is crossed by the electron beam. The most important components are parallel electrodes, which have a separation of ~1 μm and are used to produce the vortex. The remaining electrodes are used to define the boundary conditions for the projected potential that determines the phase.

The MEMS device is inserted into the microscope in a sample holder or an aperture holder that has 8 electrical contacts [19] and is connected to a generator outside the microscope, which is in turn controlled by a Labview interface. The protruding electrodes (the "chopsticks") are connected to two of the eight contacts. Thick (30 μm) electrodes are used here, resulting in a larger phase effect and therefore a larger value of OAM. The other contacts are connected together through a resistive path, so that the boundary reaches smoothly-varying potential values at different points. Such a smooth gradient would not be possible using only the 6 remaining contacts. The path includes labyrinth-like structures, which provide a sufficiently large resistance to avoid a high current that would break or melt the circuit. It should be noted that the use of controlled ohmic heating can also be used to reduce contamination during operation of the device.

The boundary conditions, which will be described in detail in a separate paper, are optimised for small OAM applications. In the present work, the boundary electrodes are grounded. The effect of grounding these electrodes is small here, since a 10 μm aperture is used to limit the field of view, as shown in Fig. 2.

For realisation of the experiment, the MEMS device was placed in the sample plane in a Thermo Fisher Titan transmission electron microscope operated at 300 kV. Large field of view off-axis electron holograms and "low magnification" diffraction patterns were then recorded with the objective lens off. Under these conditions, the objective aperture lies nearly in the same plane as the device and can be used as a limiting aperture. For off-axis electron holography, a Gatan K2 camera was used. Unfortunately, a vacuum reference wave could not be obtained from a field-free area. Instead, a minimally-perturbed vacuum reference wave was obtained from the region in front of the chopsticks, close to the grounded boundary. Simulation indicate that this still does not alter significantly the overall phase difference at the two sides of the "chopstics" and therefore the evaluation of <$\ell$>.

Figure 2 shows a reconstructed electron optical phase image recorded using off-axis electron holography for bias voltages applied to a single "chopstick" of ±2 V (*i.e.*, a voltage difference of 4 V). Based on such measurements, bias voltages of ±1 V correspond to an OAM of 67±7 $\hbar$, while bias voltages of ±2 V correspond to an OAM of 135±7 $\hbar$. As a result of distortions, accurate phase quantification could only be carried using off-axis electron holography out on a vortex corresponding to an OAM of fewer than a few hundred quanta. Furthermore, as a result of obstruction by the chopsticks, the measurements of OAM contained relatively large uncertainties.

In order to infer the OAM values for larger quanta, when off-axis electron holography could not be used, it was confirmed that the apparent vortex size in focus is linearly proportional to the chopsticks bias. This behaviour was previously shown [10] for aperture-limited vortex beams and is expected theoretically based on the stationary phase approximation. In contrast, it does not hold, for example, for Laguerre-Gauss beams, which are characterised by a beam size at focus that increases as $\sqrt{\ell}$. By combining this linear relationship with the measurement obtained using off-axis electron holography in the low bias regime, the OAM values could be extrapolated to conditions that could not be accessed holographically. Figure 3 shows experimental images of vortices, which, surprisingly, look alike apart from a scaling factor.

The maximum OAM value that could be reached in the present experiments was $\ell \approx 1067\,\hbar$ at a 16 V bias voltage . An attempt to use a higher bias voltage of 18 V unfortunately caused device failure as a result of the electrostatic pull between the two electrodes of the chopsticks shorting the circuit and leading to electrode melting. The maximum voltage that can be applied to the chopsticks depends on the elasticity of the material and on the mutual capacitance of the device. Negative feedback beyond a threshold value of deformation and enhanced attraction between the electrodes ultimately makes collapse of the structure inevitable. The formula for the critical voltage is [34]

$$U_{max} = \sqrt{\frac{8\,K_{eff}d_0^{\,3}}{27\,\varepsilon_0 A_{eff}}}, \tag{1}$$

where $d_0$ is the initial separation of the electrodes, $\varepsilon_0$ is the vacuum permittivity and $A_{eff}$ is the shared area between the electrodes, including a correction that depends on the details of the fields. In Eq. 1, the effective elastic constant is

$$K_{eff} = \frac{Ebh^3}{3l^3}, \tag{2}$$

where *E* is the Young's modulus of the material and the other geometrical parameters are described in Fig. 1. Here, $K_{eff}$ is modified by a factor 2 because the chopsticks comprises two equal electrodes facing each other. By using this formula and including a correction of 20% (see, *e.g.*, [34]) in $A_{eff}$, we obtain a value for $2U_{max}$ on the order of 33 V, *i.e.*, a bias of +16.5 and -16.5 on each electrode, in close agreement with our experimental observations. Interestingly, Eq. 1 indicates that the critical voltage is reached almost irrespective of the thickness *b* of the electrodes in the electron propagation direction. This situation becomes apparent by substituting Eq 2 and the relationship $A_{eff} \approx b\,l$ into Eq. 1. At the same time, the device thickness increases the phase scaling factor, meaning that a thicker device can be used to provide larger quanta of OAM for the same value of applied bias voltage. This has not been possible so far for fabrication limitations.

It should be noted that the linear dependence between vortex size and OAM can be explained using stationary geometrical ray tracing as a result of the location of the chopsticks in the probe forming aperture. The aperture transfer function can be written in the form $A(\bar{\rho}) = \Pi(\rho)\exp{(i\ell\theta_\rho)}$, where $\bar{\rho} = \rho, \theta_\rho$ are coordinates in the aperture plane and $\Pi_R(\rho)$ is a top hat function, which takes a value of 1 for $\rho < \rho_{max}$ and 0 otherwise. The presence of the top hat function means that the generated wavefunction is necessarily bandwidth-limited. Considering geometrical propagation from each point P in the diffraction plane, each point in the aperture plane is mapped uniquely onto a point in the probe plane $\bar{r} = \lambda f\,\overline{\nabla_{\bar{\rho}}}\varphi\big|_P$ , where $\lambda, f$ are the wavelength and focal distance, respectively. In the probe plane, the vortex beam is a circle. As it is bandwidth-limited, the azimuthal phase gradient is limited according to the expression $\frac{1}{|r|}\frac{\partial \phi}{\partial \theta_r} = \frac{\rho_{max}}{\lambda f}$ . At the same time, according to the definition of OAM, $\frac{\partial \phi}{\partial \theta_r} = \ell$. Therefore, the size of the vortex increases linearly with $\ell$. This dependence can be regarded as a limitation, as the most important parameter in the search for spin non-paraxial effects is the ratio between transverse and longitudinal momentum |p|, which is fixed. An increase in the

non-paraxial effects with $\ell$ can only be obtained if the size of the vortex is constant, independent of $\ell$. This situation can be realised more closely if the chopsticks are positioned in a different plane from that of the limiting aperture and as close as possible to the probe plane, so that the additional momentum cannot produce too large an effect on propagation. This idea will be tested in a future study.

In conclusion, a new MEMS-based device has been used to generate very large tuneable electron vortex beams in a transmission electron microscope with more than 1000 quanta of orbital angular momentum without the presence of additional beams. The generation of even larger values of orbital angular momentum is currently limited by breakdown of the device due to electrostatic forces. A strategy for increasing non-paraxial effects that can be used for studies of fundamental effects in microscopy has been presented.

**Acknowledgments**

We acknowledge the support of the European Union's Horizon 2020 Research and Innovation Programme under Grant Agreement No. 101035013 MINEON (H2020-INNOVATION LAUNCHPAD). This project has received funding from the European Union's Horizon 2020 Research and Innovation Programme (Grant No. 823717, project "ESTEEM3").

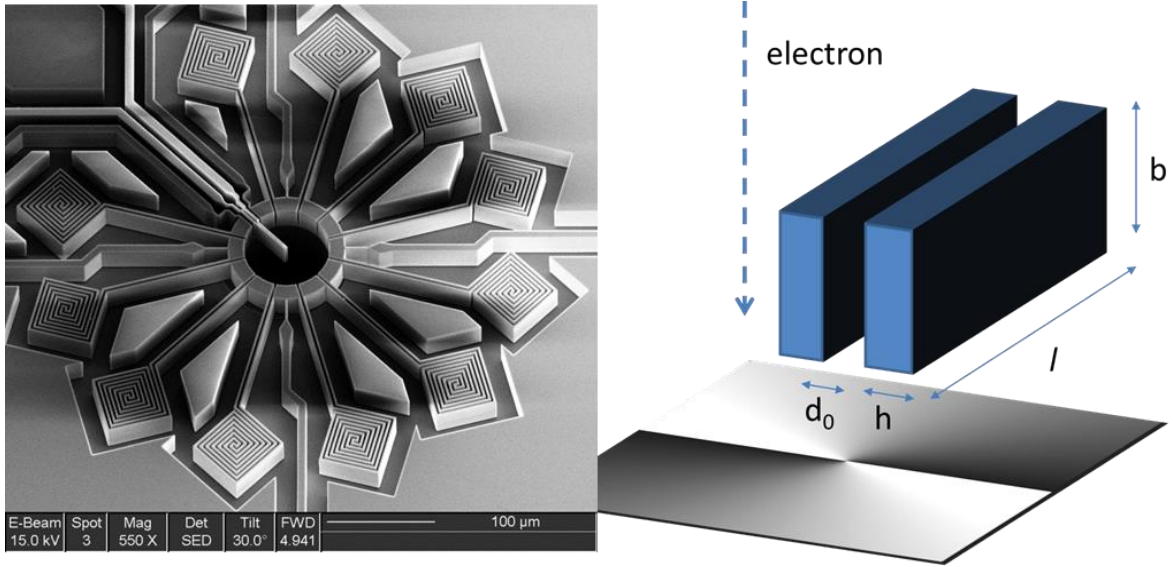

Fig. 1. (Left) SEM image of the "MINEON" MEMS device. The used of labyrinth-like structures increases the electrical resistance and allows for reasonable current flow below 1 mA. (Right) Primary geometrical parameters of the "chopsticks".

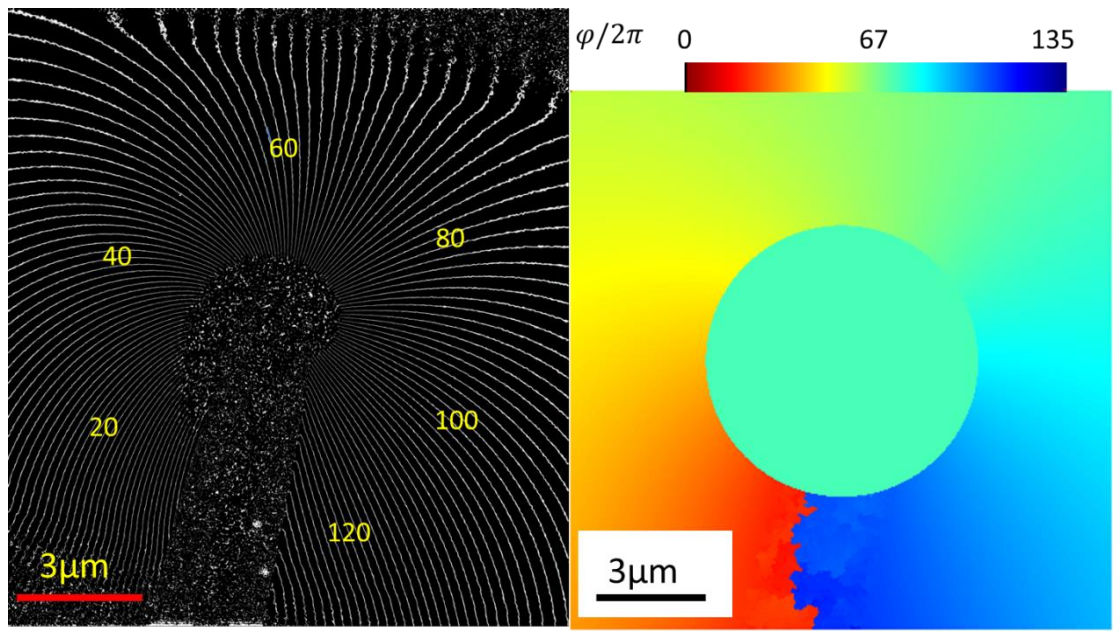

Fig. 2. (left) Off-axis electron holographic reconstruction of the electron optical phase distribution around the chopsticks with bias voltages applied to the electrodes of ±2 V. every line represents a phase increase by 2π (measured in radians). (right) Unwrapped phase of the same measurement. By extrapolation, it is possible to infer that the OAM value < $\ell$ > is approximately 137.

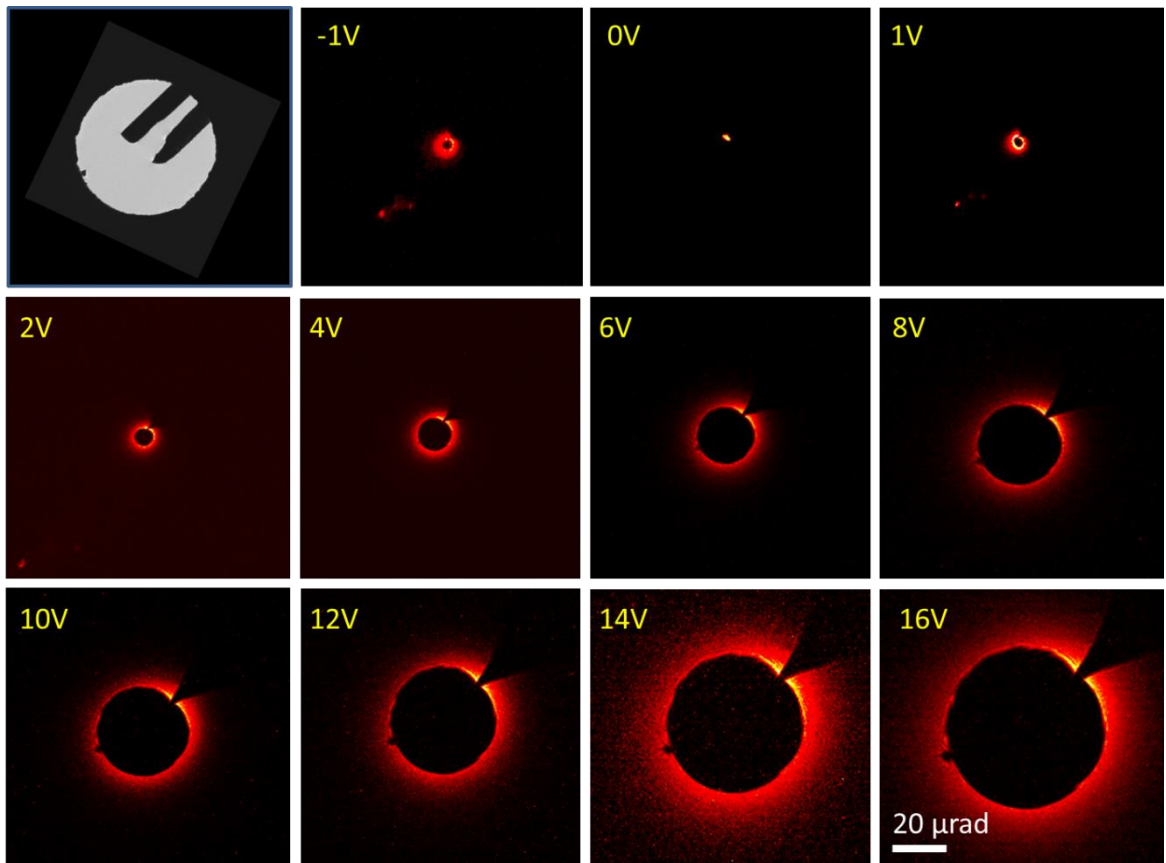

Fig. 3. Upper left: Configuration of the aperture and "chopstick" recorded in TEM mode. Subsequent images: Vortex beams (recorded in the diffraction of the aperture) shown as a function of bias voltage applied to one of the chopsticks (*i.e.*, semi-difference between the two bias voltages).

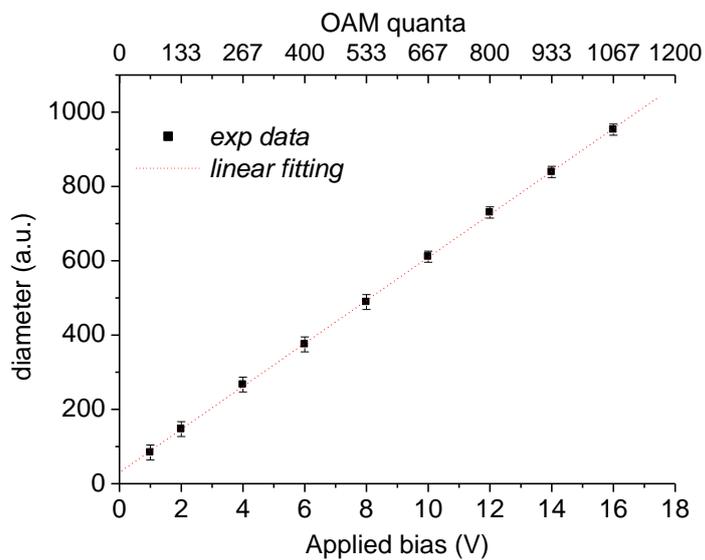

Fig. 4. Experimental measurements showing a linear relationship between the bias voltage applied to one of the chopsticks and the apparent diameter of the electron vortex beam. The upper scale shows the OAM value based on a calibration performed using off-axis electron holography.